# Dynamic high efficiency 3D meta-holography in visible range with large frames number and high frame rate based on space division multiplexing design

Hui Gao*[1], Yuxi Wang*[1], Xuhao Fan[1], Fenzhang Jiao[1], Jinsong Xia[✉1], Wei Xiong[✉1], Minghui Hong[2]

Affiliation:
1. Wuhan National Laboratory for Optoelectronics, Huazhong University of Science and Technology,1037 Luoyu Road, Wuhan, 430074, China
2. Department of Electrical and Computer Engineering, National University of Singapore, Engineering Drive 3, Singapore 117576, Singapore.

Email: jsxia@hust.edu.cn, weixiong@hust.edu.cn

*: These authors contributed equally to this work.

## 1. Introduction

As a technology that records and reconstructs wavefronts of light, holography is an ideal method for naked eye three-dimensional (3D) display, optical data storage, and optical information processing. But the traditional hologram method could not create the holographic reconstructions of virtual objects or dynamic display. To break these limitations, Brown and Lohman invented computer-generated holography (CGH) which calculating the phase map on the interference pattern by physical optics theories in 1966[1]. And by using digital devices, such as spatial light modulator (SLM) or digital micro-mirror device (DMD), CGH also could achieve dynamic holographic display[2]. However, there are still some challenges for CGH with SLM/DMD for the large pixel sizes, such as small field of view (FOV), twin images, multi-diffraction orders, et al[3].

In recent years, with the enormous development of nanofabrication technology, metamaterials and metasurfaces open new boundaries for hologram and other research fields of engineering optics[4]. Metamaterials are consisted of subwavelength artificial structure and possible to achieve novel functions beyond the limitations of bulk material. It's so difficult for the fabrication of three-dimensional metamaterials that metasurfaces play a significant role as optical devices in visible range. As a kind of two-dimensional metamaterial consisted of sub-wavelength nanostructures, metasurfaces offer us a powerful tool to achieve light modulations of amplitude, phase and polarization. It was shown that conventional optical rules should be recast into a more general form[5]. Nowadays metasurfaces have been used to achieve many different functions, such as metalens[6,7], beam splitters[8], beam deflections[9], orbit angular momentum (OAM) devices[10-12], et al.

And there are several major advantages of metasurface holography for its subwavelength unit structures, including large FOV, high resolution, elimination of high diffraction orders, and so on[3]. Meta-holography can be assorted as three classifications with different physical mechanisms, such as phase-only meta-holograms, amplitude-only holograms, and complex amplitude holograms. Most meta-holography researches in visible range are designed as static devices which could only show single frame with one piece of element. But as we know, dynamic design is essential to ideal meta-holographic smooth display. And by analogy between meta-hologram and other common dynamic display technology, we can see that there are two significations to achieve the goal. First one is frames number which means how many different frames single meta-hologram element can show. Second one is the frame rate (reciprocal of switch time between two frames) of meta-holographic display which we can quantify by factor of "frames per second (FPS)". Those discrete reconstructed holographic frames will be perceived and interpreted as a smooth video due to eye persistence. It is generally accepted that the video display would be continuous if frame rate is higher than 24 FPS. And frame rate is higher, the display effect is finer and smoother.

There are also some related research works about dynamic meta-holography as shown in Table 1 which provide an overview of the state-of-art development in visible range. Dynamic meta-holography can be divided into two classifications by different operation principles. The first classification is utilizing switchable metasurfaces which would change themselves with external controls. Many different methods are developed to realize this goal, such as using phase-change materials (e.g. $Ge_2Sb_2Te_5$, GST)[13], applying stretchable substrates[14], changing optical characteristics by chemical reactions[15], or rewriting graphene oxides metasurfaces with femtosecond lasers[16]. Another operation principle is utilizing static multiplexing metasurface. There are various multiplexing methods in recent research advancements from different experimental aspects, such as wavelength[17-19], incident angle[20], polarization[21-24]. Also, complex modulations of incident light are used as a multiplexing factor, for example, orbital angular momentum (OAM)[25]. All these dynamic meta-holography research works are meaningful and inspired. However, there are still remaining several critical challenges for dynamic meta-holography in visible range from this summary table. First, the efficiency of most works is lower than 50% (even <1% in many works) because of materials and fabrication issues. Second, as we discussed above, it is essential to display plenty of different frames with one piece of meta-hologram element to achieve a meaningful holographic video, but most current designs can only display few different frames in the experiment. Third, the modulation time of most works were too long to show smooth holographic videos. From this table, it can be seen that most research works never mention frame rate of dynamic meta-holography. The recent research work about OAM multiplexing meta-holography (Ren, H. et al. Nat. Commun., 2019) was expected to achieve relatively smooth dynamic hologram with $2^{10}$ frames number and 60 FPS[25]. But it could not achieve good display quality especially at large topological numbers conditions because of the principle limitation. In a word, there is still not such a research work about meta-holography with high efficiency and good display quality in the visible range that can show smooth holographic videos with a large frames number and high frame rate.

**Table 1** Summary of different dynamic meta-holography ("/" means no related data in the references)

| Operation principle | Methods | References | Working mode | Working Wavelength (nm) | Efficiency | Frames number | Frame rate (FPS) |
|---|---|---|---|---|---|---|---|
| Switchable metasurfaces | Phase-change material | Lee, S. Y. et al. *Sci. Rep.* (2017)[13] | Reflection | 473, 532, 660 | / | No limitation in theory | / |
| | Stretchable substrate | Malek, S. C. et al. *Nano. Lett.* (2017)[14] | Transmission | 633 | / | 6 | / |
| | Chemical reactions | Li J. X. et al. *Sci. Adv.* (2018)[15] | Reflection | 633 | / | 10 | 1/40 |

| | | | | | | |
|---|---|---|---|---|---|---|
| | Rewriting metasurface | Li X. et al. *Nat. Commun.* (2015)[16] | Transmission | 405, 532, 632 | 15.88% | No limitation in theory | / |
| Multiplexed metasurfaces | Wavelength division multiplexing | Li X. et al. *Sci. Adv.* (2016)[17] | Transmission | 380-780 | 23.7% | 7 | / |
| | | Wang B. et al. *Nano. Lett.* (2016)[18] | Transmission | 473, 532, 633 | 18% | 3 | / |
| | | Huang Y. M. et al. *Nano. Lett.* (2015)[19] | Reflection | 405, 532 658 | 0.3% | 3 | / |
| | Angle division multiplexing | Zhang X. et al. *Nanoscale* (2017)[20] | Transmission | 405 | 0.3% | 8 | / |
| | Polarization multiplexing | Chen, W. et al. *Nano. Lett.* (2014)[21] | Reflection | 405, 633 780 | 18% | 3 | / |
| | | Zhao R. et al. *Light Sci. Appl.* (2018)[22] | Transmission | 800 | 15.97% | 7 | / |
| | | Mueller, J. P. B. et al. *Phys. Rev. Lett.* (2017)[23] | Transmission | 532 | / | 2 | / |
| | | Wen, D. et al. *Nat. Commun.* (2015)[24] | Reflection | 475-1100 | 59.2% | 5 | / |
| | Orbital angular momentum | Ren, H. et al. *Nat. Commun.* (2019)[25] | Transmission | 633 | 65% | $2^{10}$ | 60 |
| | Space division multiplexing | Current work in this paper | Transmission | 633 | 70.2% | $2^{28}$ | 9523 |

In current work, we demonstrate a new design of meta-holography in visible range based on space division multiplexing metasurface which can achieve $2^{28}$ different holographic frames and very high frame rate (maximum of frame rate, 9523 FPS). Also, the metasurface is consisted of silicon nitride ($SiN_x$) nanopillars with high efficiency (more than 70%).

**2. Design and realization of dynamic space division multiplexing meta-hologram**

To achieve our goal, we can get some inspiration from the analogy between dynamic meta-hologram and common two-dimensional (2D) display technologies in our daily life. As we all know that the ideal solution means of dynamic meta-holography is perfectly controlling each nanostructure of metasurface. It means we need to control each pixel of the element independently in high speed, just like how LED or LCD screens work. However, it's beyond the limitation of our technology nowadays to do similar things to metasurface in visible range. But except these pixel display screens, there are also two kinds of other methods to achieve dynamic 2D display. One is dividing the whole graph into many different subgraphs and combining different subgraphs together at different time, e.g. digital tubes display on the electronic scoreboard or electronic meter. Another one is displaying different frames from a continuous video at different time, e.g. conventional movies recorded and projected by cinefilms. It can be concluded that both of them are space multiplexing methods.

The physical mechanism of space division multiplexing meta-hologram is demonstrated and illustrated in Fig. 1. Traditional meta-hologram design was calculating the corresponding phase map or amplitude map of target reconstructed objects with mathematical algorithm (e.g. GS algorithm) in whole fabricated metasurface region. To create space division multiplexing meta-hologram, the metasurface would be divided into N different subregions. And the corresponding target reconstructed object for every independent region is different but associated with each other. Different target reconstructed objects are either subgraphs of a whole holographic graph or continuous frames of a holographic video. As marked by different colors in Fig. 1(a), *nth* subregion would reconstruct corresponding *nth* object. By modulating the space distribution of incident light to illuminate different subregions at different time, the reconstructed frames would also change along time. To be noted that reconstructed frames can be 2D pictures as well as 3D objects.

$$E(x, y, t) = \left[\sum_1^N F_n(t) * A(x_n, y_n)\right] * \left[\sum_1^N e^{-i*\varphi_n(x_n, y_n)}\right] \quad (1)$$

Eq. (1) shows the basic idea of dynamic space division multiplexing meta-hologram design. *E(x,y,t)* means the modulated complex amplitude by designed metasurface at time *t*. The phase map of static metasurface is

$$\phi(x, y) = \sum_1^N \varphi_n(x_n, y_n), \quad (2)$$

which is consisted of N different regions with different corresponding reconstructed target objects, and *A* means the equal amplitude of incident planar beam, *F(t)* is the binary function which can only equals to 1 or 0. So

$$A(x, y, t) = \sum_1^N F_n(t) * A(x_n, y_n) \quad (3)$$

means the modulated light distribution of incident beam at time *t*. The number of different frames that this design can display is $2^N$.

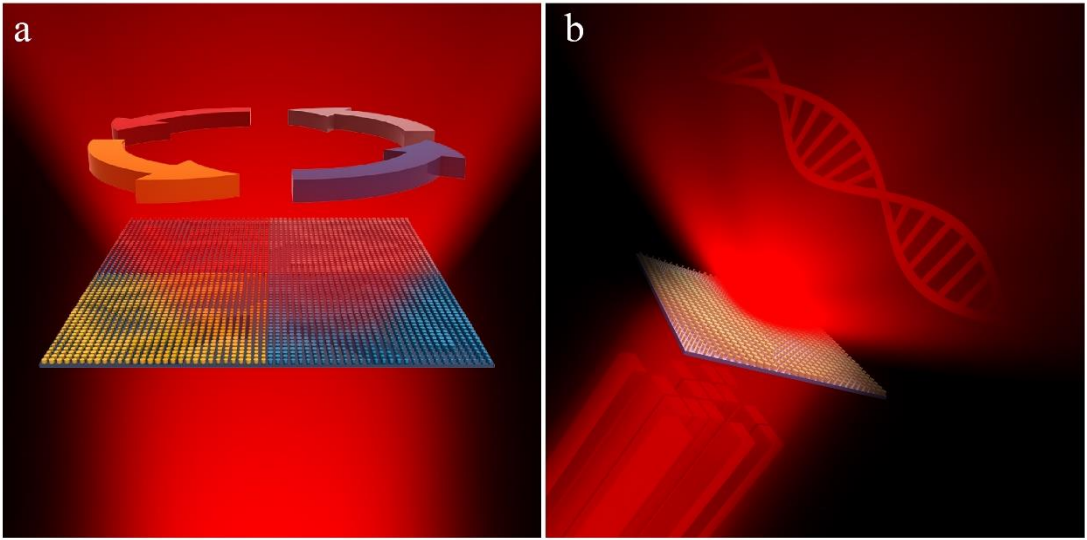

**Fig 1** The schematic diagram of space division multiplexing meta-hologram design. **a.** The design of space division multiplexing metasurface. Different regions (marked as different colors) reconstruct different but associated target objects. **b.** Sketch map of dynamic meta-holographic display. By coding and modulating the incident beam to illuminate different combined regions of metasurface, the system can display dynamic holographic videos.

It can be concluded from the equations above that there are two significant points to achieve our dynamic meta-holography design. First is dynamic beam modulation part to code the space distribution of incident beam. The module can be achieved by the projection system composed of DMD, lens and microscope objective as shown in Fig. 2(a). The incident light is modulated by DMD in high speed, e.g. maximum 9523 Hz in our experiment. The lens and microscope objective perform as a 4f system to narrow the coded incident beam to illuminate different regions of metasurface. Another essential part is static space division multiplexing metasurface with high efficiency in visible range. In this work, the static metasurface is consisted of $SiN_x$ nanopillars as shown in illustration of Fig. 2(b). The absorption coefficient of $SiN_x$ material is so little that $SiN_x$ is nearly transparent in visible range. And its refractive index n is near 2 which is far larger than normal glass materials. These characteristics make $SiN_x$ material be suitable to design high efficiency metasurface with equivalent refractive index method in visible range. The heights of $SiN_x$ nanopillars are all same with 700nm and periods of rectangle lattice are 500nm while the radii vary from 90nm to 188nm. The nanopillars were simulated by FDTD and we choose six proper radii for fabrication. The characterization of amplitude transmission efficiency and phase response of $SiN_x$ nanopillars are shown in Fig. 2(b) as a function of nanopillars radius at a wavelength of 632.8nm, respectively. It indicates that the transmission efficiency keeps almost constant and very high (more than 95%) while phase response varies from 0 to $2\pi$. Fig. 2(c) presents the fabricated results with optical image (left) and scanning electron microscope (SEM) images (right). The scale bar in optical image and SEM images are 100μm and 1μm, respectively.

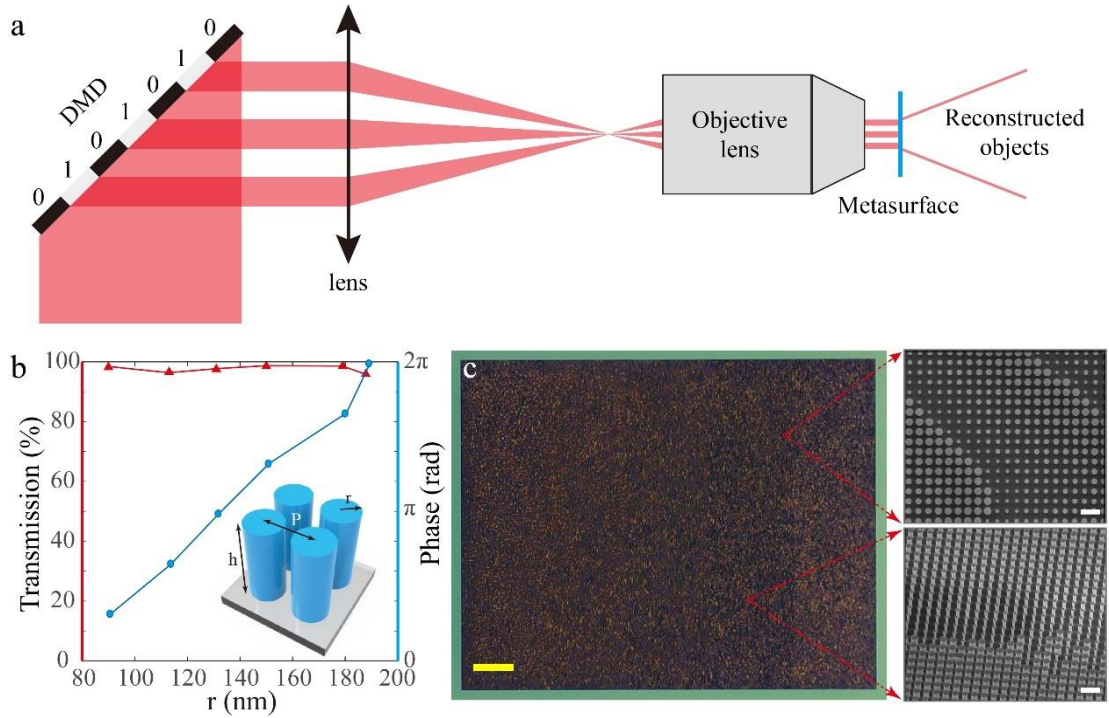

**Fig. 2** The realization of dynamic space division multiplexing meta-hologram. **a.** Dynamic space beam coding module. DMD modulate the incident light in high speed, e.g. maximum 9523 Hz in our experiment. The lens and microscope objective perform as a 4f system to narrow the coded incident beam to illuminate different regions of metasurface. **b.** The geometrical diagram of SiN nanopillars. **c.** The characterization of amplitude transmission efficiency and phase response of SiN

nanopillars as a function of nanopillars radius at a wavelength of 632.8nm, respectively. And the illustration is geometrical diagram of SiN nanopillars. **c.** The fabricated results with optical image (left) and scanning electron microscope (SEM) images (right). The scale bar in optical image and SEM images are 100μm and 1μm, respectively.

## 3. Experiment results
### 3.1 Dynamic combined subgraphs meta-hologram

As we have discussed above, one of design methods for dynamic display is dividing whole picture into subgraphs and showing different frames by combination of different subgraphs. This method also can be used in the design of dynamic combined subgraphs meta-hologram (CSMH). In this paper the metasurface holographic digital tubes display system is designed to demonstrate the method as shown in Fig. 3(a). The whole target reconstructed image is the digital tube pattern of "88:88" which is consisted of 28 subgraphs. Accordingly, the metasurface is divided into 28 different subregions which reconstructed corresponding subgraphs as marked by numbers in Fig. 3(a). And the example of frame "12:12" is demonstrated in figure. The size of each subregion is 150*200 μm$^2$ and whole size of fabricated metasurface region is 1050*800 μm$^2$. By coding the space distribution of incident beam, metasurface can reconstruct tremendous number of different frames. To some extent it was a kind of share aperture design. Assuming the number of subregions was N, the total amount of different frames was $2^N$ because the binary function *F(t)* could be 0 or 1 for each subregion. For example, N is 28 in this design, so the total frames number is $2^{28}$= 268 435 456.

And, of course, it's impossible to show each different frame in this paper. So, ten typical frames are presented in Fig. 3(b) to demonstrate the design. The illustrations in first and third rows exhibit the experimental results which varied from 00:00 to 99:99. In the meantime, second and fourth row show corresponding coding pattern of DMD. To be noted that there are dark gaps between coding patterns of DMD as shown in illustrations while no gaps between subregions of fabricated metasurface. There is a little divergence for narrowed laser beams to illuminate subregions because of geometrical optical aberrations and diffraction. Dark gaps could restrain crosstalk of adjacent subregions. The frame rate of meta-holographic digital tube displaying was depended on the switch time of coding pattern of DMD. In our experiment, the minimum switch time of DMD is 105μs thus the frame rate could vary from 0 to 9523 FPS which was far more than the limitation of persistence of vision. It is demonstrated that the design could achieve smoothly dynamic meta-holographic display (see Supplementary Video 1).

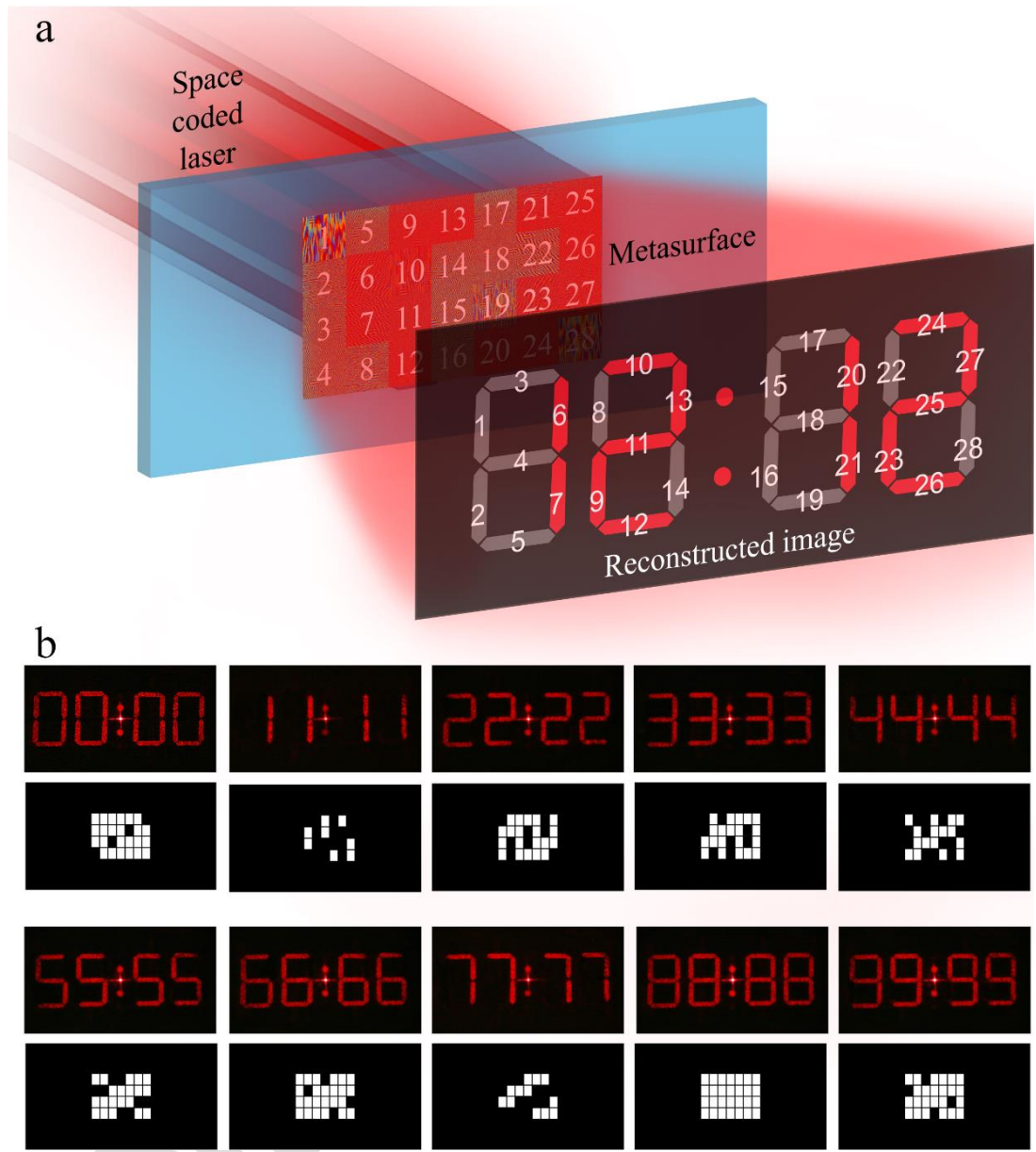

**Fig. 3** The design and experiment results of dynamic CSMH. **a.** The whole target reconstructed image was the digital tube pattern of "88:88" which is consisted of 28 subgraphs. Accordingly, the metasurface was divided into 28 different subregions which reconstructed corresponding subgraphs as marked by numbers. The example of frame "12:12" was demonstrated in figure. **b.** The results of ten typical frames which varying from 00:00 to 99:99. The illustrations in first and third rows were experiment results while second and fourth row showing corresponding coding pattern of DMD (see Supplementary Video 1).

### 3.2 Dynamic continuous frames meta-hologram

Another design in this paper is dynamic continuous frames meta-hologram (CFMH) which is similar to conventional movies recorded and projected by cinefilms. The metasurface sample is divided into many subregions which would reconstruct different frames from a continuous video. In this design, twenty continuous frames from a short video which showing the rotation of four capital letter "HUST" are selected as the reconstructed frames of dynamic meta-hologram as shown in Fig. 4. The incident laser beam is modulated by DMD as space scanning beam and illuminates different single subregion of metasurface in designed sequence. Then the reconstructed frames would change along time to display the dynamic meta-holographic movie and similarly the frame rate of holographic video is depended on the switch time of DMD. The experiment results for each frame are presented in Fig. 4(b). The short meta-holographic video demonstrate the practicability of this design method (see Supplementary Video 2).

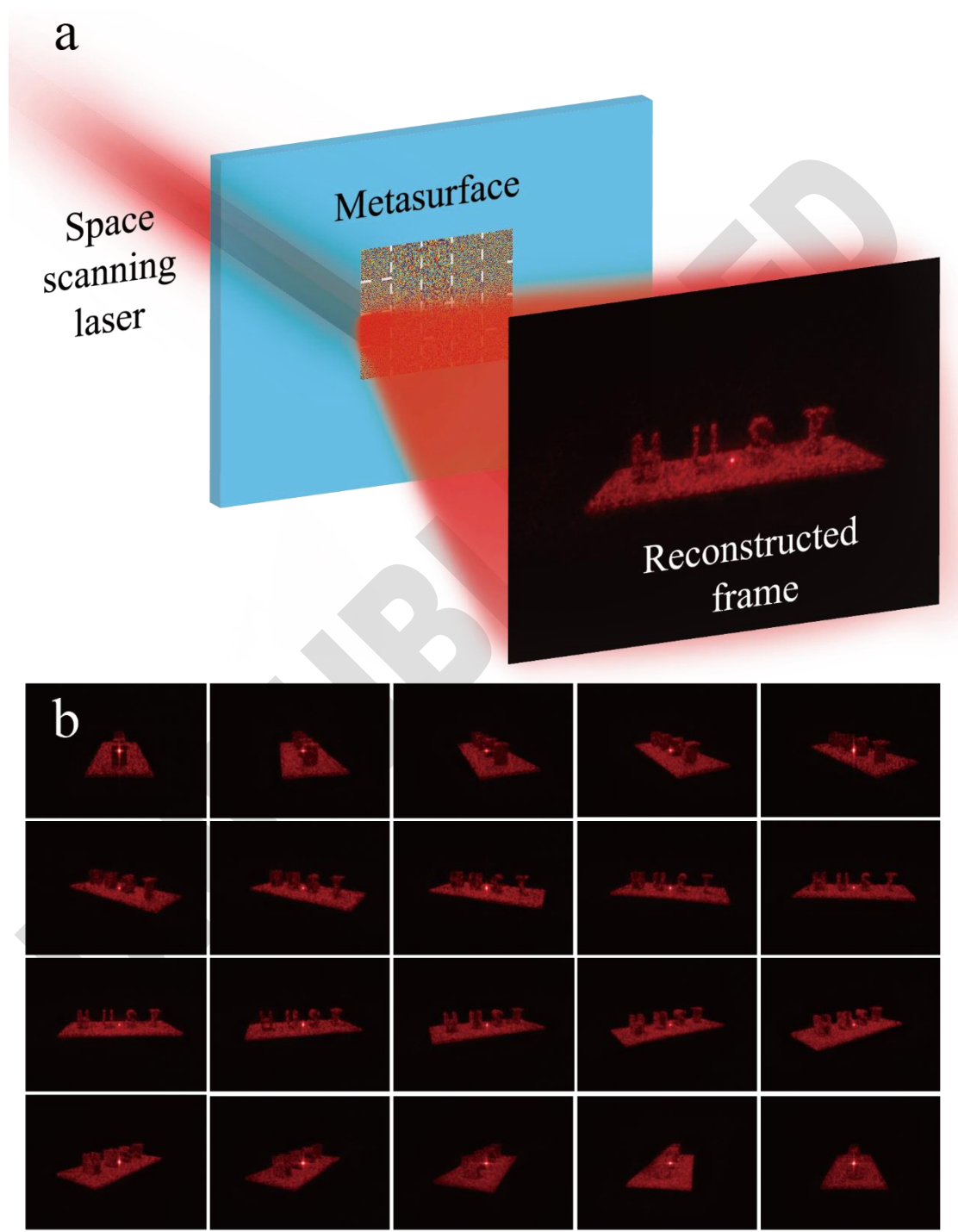

**Fig. 4** The design and experiment results of dynamic CFMH. **a.** The schematic diagram of design. The metasurface sample was divided into subregions which would reconstruct different frames from a continuous video. The incident laser beam was modulated by DMD as space scanning beam and illuminated different single subregion of metasurface in designed sequence. **b.** Experimental reconstructed results for each different subregion of metasurface.

### 3.3 Dynamic 3D meta-hologram design

CFMH design can be used to display 2D as well as 3D holographic videos. A metasurface is designed to show 3D holographic video in this paper as shown in Fig. 5(a). The whole annular meta-holographic element is divided into eight subregions, and each region is designed to reconstruct a 3D arrow in free space. The geometrical parameters are marked in Fig. 5(a), the internal radius of annular metasurface is r=150μm while outer radius R=450μm. The reconstructed 3D arrows are designed in the central circle with radius $r_0$=125μm and height between $h_1$=2000μm and $h_2$=2020μm. Eight 3D arrows are positioned end to end in free space. The reconstructed light field of each 3D arrow is detected by home-build microscope along z-axis as shown in Fig. 5(b). It demonstrates the design method can be used in smooth meta-holographic display (see Supplementary Video 3).

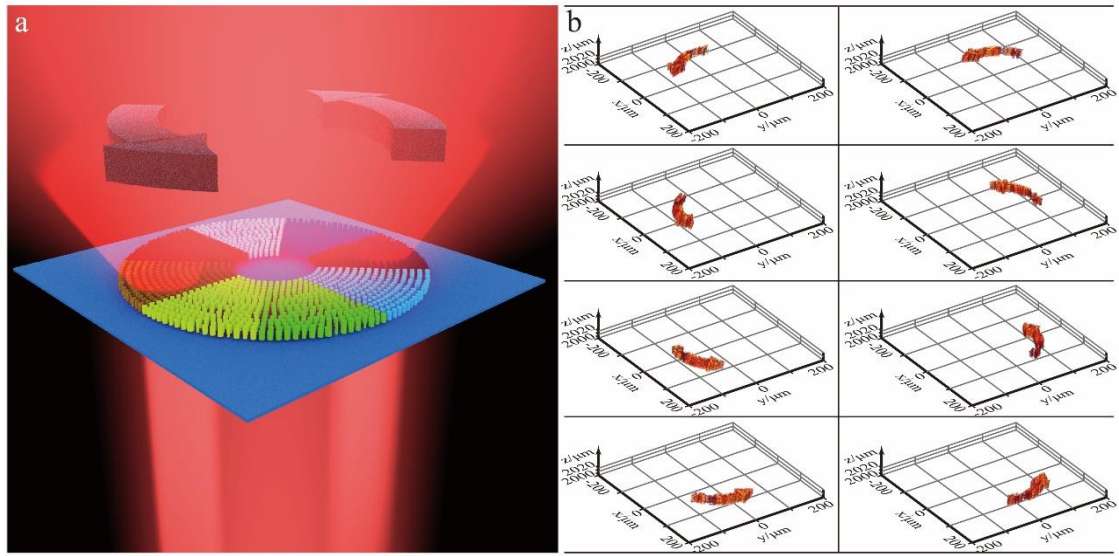

**Fig. 5** The design and experiment results of dynamic 3D CFMH. **a.** The schematic diagram of design. The annular metasurface sample was divided into eight subregions which would reconstruct a 3D arrow in free space (the diagram is not in real scale). **b.** Reconstructed light fields of 3D arrows for different subregion of metasurface.

### 4. Discussion

Our demonstration provides a new method to achieve dynamic meta-holography in visible range based on dynamic space coding of incident beam by DMD and space division multiplexing metasurface design. The fabricated metasurfaces are divided into N different subregions and the reconstructed holographic objects are different with each other. Three designs are demonstrated in this paper. First one is combined subgraphs meta-hologram which could display $2^N$ different frames (more than two hundred million in this paper). Another one is continuous frames meta-hologram which could show complex holographic videos. And third one is also a CFMH design to show 3D holographic videos. Both three designs could achieve high speed frame rate (0~9523 FPS) which is far more than the limitation of visual residue, so that the designs in this paper display fine and smooth holographic videos. The metasurface is consisted of $SiN_x$ nanopillars which are designed by equivalent refractive index method and simulated by FDTD. It should be noted that the period of nanopillars was larger than λ/n (316nm) which resulted little higher order diffraction. But the high order diffraction efficiency is so low that would not influence the experiment results. Notably, the reported dynamic meta-holograms feature a high efficiency (more than 70%) in the visible range. The large frames number, high frame rate and high efficiency mean this method can not only satisfy the needs of holographic display but also be suitable for many different research fields, such as laser fabrication, photolithography, 3D forming of two-photon polymerization and optics information processing, and so on.

5. Method

**Fabrication of $SiN_x$ metasurface.** The fabrication of $SiN_x$ metasurface was began from a glass wafer with a thickness of 500 μm as the substrate. A silicon nitride layer ($SiN_x$, n=2.023 when λ is 633 nm) with the thickness of 700nm is deposited by plasma enhanced chemical vapor deposition (PECVD) on the substrate. Then, a chromium (Cr) layer with the thickness of 20 nm is evaporated by electron beam evaporation (EBE) on the top of $SiN_x$ layer as a hard mask. Next, a photoresist layer (CSAR62) of 200 nm is spin-coated onto the top of Cr layer. The Hologram pattern is written by electron-beam lithography (EBL, Vistec: EBPG 5000 Plus), and defined into the photoresist layer after development and fixation. Afterwards, the hologram pattern is transferred into the Cr hard mask layer by inductively-coupled plasma etching (ICP, Oxford Plasmalab: System100-ICP-180), and the residual photoresist is stripped by oxygen plasma stripper (Diener electronic: PICO plasma stripper). Finally, the pattern is transferred into the $SiN_x$ layer by another ICP process, and the remaining Cr is removed by Cr corrosion solution. The reason for our utilization of Cr layer as hard mask is because the extremely high etching selectivity between Cr and $SiN_x$, and there is no electron-beam photoresist available (at 200 nm thickness) to provide enough selectivity versus $SiN_x$ to achieve direct ICP with 700 nm depth. While the thickness of electron-beam resist is limited by the feature size around 80nm. As a result, a two-step ICP process with a hard mask is necessary for the fabrication. (see detailed process flow chart from Supplementary Fig. 1)

**Optical setup.** The optical components and setup of dynamic space division multiplexing metasurface are shown in Fig. 6. The HeNe laser (Pacific Lasertec, 25-LHP-991-230) at wavelength of 632.8nm propagates through a spatial pinhole filter and collimating lens and becomes expanding laser beam with good beam quality. Then the expanded laser beam is modulated by DMD (Texas Instrument，DLP6500FYE) in high speed. The coded beam propagates through the 4f system consisted of lens and microscope objective and the narrowed beam illuminates particular regions of metasurface. The reconstructed holographic frames are collected by Fourier lens or objective lens and recorded by CCD.

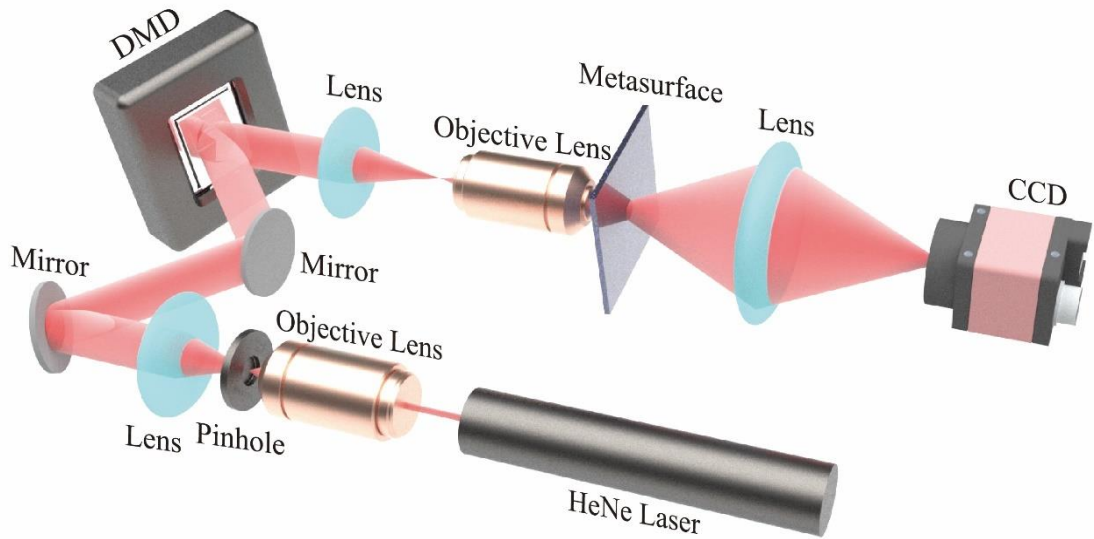

**Fig. 6** The schematic diagram of optical setup. The HeNe laser at wavelength of 632.8nm propagates through a spatial pinhole filter and collimating lens and becomes expanding laser beam with good beam quality. Then the expanded laser beam is modulated by DMD in high speed. The coded beam propagates through the 4f system consisted of lens and microscope objective and the narrowed beam illuminates particular regions of metasurface. The reconstructed holographic frames are collected by Fourier lens or objective lens and recorded by CCD.